\documentclass{article}
\usepackage[utf8]{inputenc}
\usepackage{graphicx}
\usepackage{lipsum}

\title{Security Monitoring System Using FaceNet For Wireless Sensor Networks}
\author{Preetha S. (preetha.ise@bmsce.ac.in) \and Sheela S. V. (ssv.ise@bmsce.ac.in )}

\date{October 2021}
\begin{document}

\maketitle
\begin{abstract}
Face Recognition techniques have been rising rapidly over past few decades and gaining attraction in many security systems. Several methodologies are proposed to provide excellent results in face detection. Sensing technology enables several applications such as transportation, healthcare and security. Video surveillance systems trace and identify moving objects. Wireless Sensor networks are used to monitor remote areas and can be applied to monitor a facility by considering each camera as sensor nodes. Cameras are used as nodes in a wireless sensor network with a central server or a gateway node for all the monitoring and analysis of the information retrieved from the nodes.  Identification and authentication of users in any organization is quite difficult due to high movement. Face recognition can be used detect faces and identify them continuously in a video feed which can be deployed to continuously monitor an area. Feeding from camera to base station uses Multi-task Cascaded Convolutional Neural Networks (MCTNN) and FaceNet algorithms for face recognition. Further information about the person is sent to all the end-user nodes present in the wireless network. This approach has been implemented and evaluated on a prototype wired camera network called FaceNet to monitor a network and provide security. A method for tracking people in 2D world coordinates and acquiring canonical frontal face images that fits the sensor network paradigm. The approach evaluates and demonstrates the tasking algorithm in action on data acquired from the FaceNet camera network. 

\end{abstract}

\section{Introduction}

The motives stem from the requirement of programmed identifications, the need for surveillance systems, human-computer interface design and the curiosity in the visual capacity of humans for face recognition has been studied. The knowledge from multiple disciplines and their research are involved in studies such as computer vision specialists, neurologists, psychologists, pattern recognition, machine learning and image processing experts. Several studies are done to pull through many challenging factors like lighting, face mood, expressions, size and pose of the face. The need for a robust model with practical cases of the everyday lives with variety of dynamics still exists.

Currently, the most utilized method of a biometric authentication is fingerprint scanning and it is being used for user identification and authentication in many places such as office security authentication, college attendance monitoring system and mobile authentication. Undoubtedly fingerprint scanners are one of the best ways to identify users with high accuracy but the disadvantages are that fingers with dirt, damage, gloves or grease affects its recognition and the possibility of somebody else’s fingerprint  is used to unlock your phone is unlikely. These drawbacks could be overcome with the face recognition as the metric for authentication. 

Face recognition is one among the most fascinating and significant subject in the research fields over the last two decades. Frameworks with added highlights empowers mechanized checking, along with making it simple to work with and a reliable application. They can occur consistently on most recognition gadgets. It is quick, easy and productive and has been a forward leap in biometric verification, also witnessed to be the most accepted form of general biometric authentication for many devices like mobile phone and laptops. In general, face recognition is owing to simplicity of use than manual input of passcodes. It is predominant in each gadget and are more straightforward than manual input of any data for unlocking the device. The probability of a random person able to unlock your phone using face is a rare situation. 

Wireless Sensor Networks (WSNs) can be regarded as a self-arranged and framework-less isolated systems to monitor environmental state. These includes sound, temperature, weight, vibration, motion or change in environmental, chemical or physical composition with sensor nodes. WSN comprises of a lot of associated modest sensor hubs, which speak with one another and trade data and information. Data from the sensor nodes will be sent to a base station or sink where the data accumulated can be stored, monitored or analyzed within a network\cite{1}. The base station or the sink acts as an interface to clients within the network system. Such networks comprises of several sensor nodes and is used in diverse applications of defence which includes tracking and sensing, environmental activities, disaster management, monitoring etc. Wireless sensor networks are used with face recognition software to monitor a secured environment.

\subsection {Application Areas}
\begin{itemize}
    \item  \emph{Security authentication}: Everyone entering the premises of an establishment, their identity can be authenticated by getting an image by a CCTV and a facial recognition system can be used on that image to verify if their entry is allowed to the premises.
    \item \emph{College attendance}: Instead of traditional name/roll call attendance system which takes effort and time with lower efficiency because of the chances of proxy responses, a face recognition system using a camera can be used to verify presence of a student in a class saving time, effort and does a more efficient job.
    \item \emph{Mobile and Social Media Account Authentication}: The front camera of the phone can be used to detect and identify people by unique features of their face.
    \item \emph{Tracking people}: A particular person can be detected by either giving the name or a photo. Then a facial recognition system can be run to get that person’s location and a tracking system can then get the footprint.
\end{itemize}

\section{Literature Survey}

Various studies have proved that biometric recognition systems provide security to environment and applications. Weiguo wan et.al proposed an approach to map feature vectors to a face image based on FaceNet \cite{2}. The experiment was conducted on public dataset which showed considerable good results. Initially photo images are assigned to gallery in sketch style without matching directly with photo images. Hence consistency is well maintained between face sketches and photos. Surveillance systems face many challenges regardless of current advances in the arena of tracking objects or patrolling through CCTV footage or any other portable surveillance object like drones. Such systems are prone to human errors. Jose Edwin et.al experimented an economical surveillance system. The system used a smart multi-camera to recognize faces based on MTCNN algorithm on Jetson TX2 and FaceNet. This system logged the presence and tracked the subject with multiple cameras to reduce errors. Implementation on the multi-camera surveillance system was focused to achieve higher accuracy in recognition \cite{3,4}.

Present methods for face recognition poses substantial challenges when implemented on a large scale. Florian Schroff et al. \cite{5} suggested a FaceNet system which studied a mapping from face images directly to a Euclidean space to measure corresponding face similarity with distances. Generated face space undergoes verification and clustering process. Feature vectors using standard techniques helps FaceNet embedding’s to be implemented easily during clustering. The method used trained deep convolutional network to optimize embedding. Tracking and identity management consider frontal face features as the most appreciated features since they do not change significantly in appearance. A technique to track people in 2D world coordinates was discussed in \cite{6}. The technique was assessed using FaceNet; a wired camera network. The method demonstrated how moving objects trajectories are sensed and utilized to achieve greater quality canonical views while conserving node energy. In \cite{7} the proposed technique estimated model parameters (length, focal, pose) inside a coherent probabilistic formulation, object tracks and 2D/3D temporal trajectories within the camera referenced system

Zuheng Ming et.al\cite{8}discussed  how FaceNet has outshined human level performance on accuracy for “Labelled Faces in the wild” and “YouTube Faces in the wild(YTF)” datasets. FaceNet practices triplet loss to recognize success for face verification. Triplets do explode when training large datasets. The proposed method presents a simple class wise triplet loss to reduce the number of probable triplets to be learned based on the intra/inter-class distance metric learning. In \cite{9} PCA (Principal Component Analysis) and LDA (Linear Discriminant Analysis) based Face acknowledgment framework was presented. The method includes two stages; first includes anticipation of face picture from the first vector space to a face subspace by means of PCA, second using LDA to get a best direct classifier. Association of PCA and LDA enhances the assumption capability of LDA.

A novel face discovery and acknowledgement framework was suggested by M Shujah Islam Sameem, et.al\cite{10}. The method had the ability to perceive human faces in single, similar to continuous numerous face pictures in a database. Pre-preparing of the proposed outline work incorporates commotion expulsion and gap filling in shading pictures. After pre-handling, face identification is performed by utilizing viola jones calculation. Recognized countenances are trimmed out of the information picture to make quick calculation. Speeded Up Robust Feature (SURF) highlights are extracted from the trimmed picture. For face coordinating, putative component coordinating is completed and anomalies are expelled utilizing M-Estimator Sample Consensus (MSAC) calculation. Single and multiple color images from class people of Graz 01 dataset are utilized to assess the framework.

Face identification and acknowledgment benchmarks have moved to progressively troublesome conditions. The test introduced in \cite{11} tends to follow the stage toward programmed discovery and distinguishing proof of individuals from open air observation cameras. While face location has demonstrated wonderful accomplishment in pictures gathered from the web, observation cameras incorporate increasingly different impediments presents climate conditions and picture obscure. In spite of the fact that face check or shut set face ID have outperformed human abilities on some datasets. Open-set recognizable proof is substantially more mind boggling as it needs to dismiss both unclear characters and fake acknowledges from the face identifier. A smart and secured attendance system was proposed in \cite{12}. The method was built on networked surveillance video. The system was designed fusing face recognition, video image processing and deep learning and integrated attendance and security functions. A person’s identity was identified by using the proposed Sliding Average Method. The experimental results verified the effectiveness of the method with false accept rate (FAR)-2.52

A novel Deformable Face Net (DFN) was used to handle the pose variations in face recognition for large dataset. The convolution module tried to concurrently study identity-preserving feature extraction and face recognition oriented alignment. Since faces pose strong structure, the proposed Displacement Consistency Loss (DCL) imposes the learnt displacement arenas to align faces to be reliable in amplitude and orientation. Detecting multiple faces from a single frame and poor resolution is challenging. The proposed method in \cite{13} combined Support Vector Machine (SVM) and FaceNet to enhance accuracy of extracted features. Features are extracted by embedding 128 dimensions per face using FaceNet and SVM to classify extracted features with training data. 

Labeled faces was used in wild benchmark, FaceNet method evaluated a machine learning technique called SVM to classify the generated embeddings to achieve accuracy \cite{14}.  FaceNet, along 5 fold cross-validations were used in \cite{15} to obtain higher and perfect accuracy, also proved to be better when compared to Openface. FaceNet algorithm combined with K-Nearest Neighbour enhanced accuracy of extracted features. The method classified the features into three classes namely fatigue, focused and unfocused \cite{16}. FaceNet model is a profound convolutional arrangement that utilizes triplet misfortune work. Triplet misfortune work, limits the separation between a positive and a grapple while expanding the separation between the stay and a negative. These are generally adjusted coordinating or non-coordinating face patches used to train the data; It limits the space between Associate in Nursing grapple and a positive, every one of that have a proportionate character, and boosts the space between the stay and a negative of a distinct identity. Triplet Loss minimizes the distance between a positive and an anchor with similar identity and maximizes the distance between a negative and the anchor of dissimilar identity.

Researchers experimented with a lot of dimensionalities and chose 128-D for the FaceNet, as it was the best performing method. It was estimated that the greater dimensionalities would perform better, but it could also mean that they need more training. During training a 128-D, float vector is used which is quantized to 128-byte vector without loss of accuracy. Smaller embedding dimensions could be employed on mobile devices, with minor loss of accuracy. The 128 dimensional embedding returned by the FaceNet model can be used to cluster faces effectively. Once such a vector space (embedding) is created, tasks such as face recognition, verification and clustering can be easily implemented using standard techniques with FaceNet embedding’s as feature vectors. In a way, distance would be closer for similar faces and further away for non-similar faces.

\section{Proposed Method}

\emph{Algorithm for Face Recognition}: FaceNet technology uses image to transform face into 128D Euclidian space similar to word embedding.  FaceNet model is trained with triplet loss for different class of faces to determine the similarities and differences among them. Face images can be effectively clustered by utilizing the 128D embedding returned by the FaceNet model. With the vector space created consisting of embedding in place clustering, verification and face recognition tasks can be easily implemented using standard techniques since the distance will be closer for similar faces and farther for dissimilar faces.

Embedding can be created when the model is trained for the face by serving the face to the model. To equate two pictures, embedding for both face photos is to be created separately after which the above formula to find the distance of the least value for like faces and highest value for dissimilar face.

\emph{Approach} :The methodology includes the following steps 
\begin{itemize}
\item Align faces using the face detection algorithms 
\item Detect faces using face net
\item Analyze the detected face data
\item Authenticate, Blacklist or whitelist the detected faces
\item Use Wireless sensor networks to communicate with monitoring systems or devices to take action on the information received
\end{itemize}

\section{Proposed Architecture }
The proposed architecture of face recognition system is shown in figure 1. A camera sensor node is connected to the base station using a wifi network. Wifi is used as a common medium between nodes and base station (server). All the connections between nodes and base stations is done through socket programming. Camera nodes send its input to the base station, this runs the facial recognition scan and then further sends this information to all end-user nodes which are connected to the network.

\begin{figure}[h]
       \centering
       \includegraphics[width=0.8\textwidth]{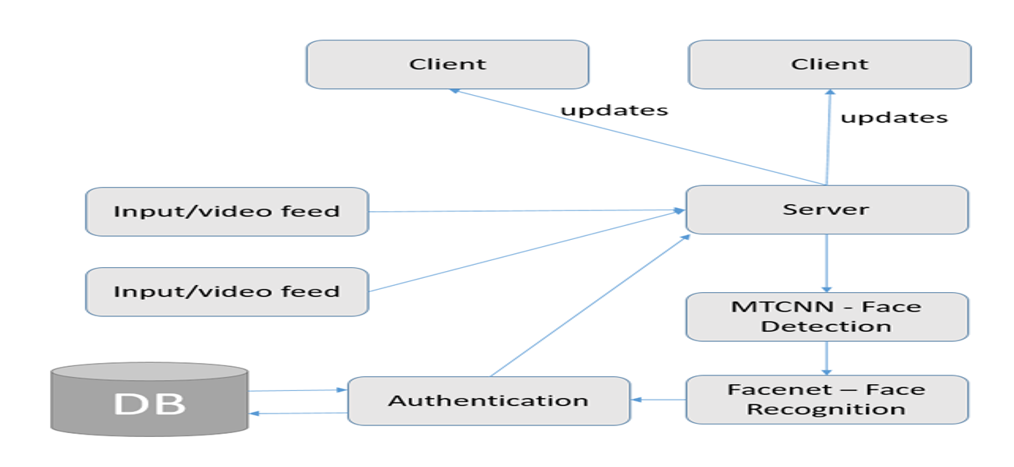}
       \caption{Proposed Architecture for face recognition}
\end{figure}

\emph{Face Detection}: The primary and essential process for face recognition to initiate is to detect a face. To determine the regions in an image, it needs to be matched or recognized against the dataset. Face detection are often considered a particular instance of detecting an object-class. Detection of object-class seeks out sizes and locations of all objects in a picture that belong to a given category. Algorithms specialize in recognizing frontal faces of human similar to detecting an image within which the picture of an individual is associated part by part. Matching method is invalidated if any changes exist within facial feature information.

Discovering and placing a face in an image discusses determining the coordinates corresponding to the face in the photograph, wherein localization is to establish the scope of the face, generally with bounding container surrounding the face. Identifying faces in a picture is effortlessly done by humans, even though it is generally difficult for machines considering the range types and quality of faces. For instance, regardless of the orientation and angles of the faces, they must be detected. More challenges are presented by conditions like illumination, dressing, accessories, make-up, hair, facial hair styles, color and age.

MTCNN and YOLO are the two algorithms used for face detection to produce excellent results in face detection, MTCNN goes along the lines of detecting the face with the help of face features and bounds the boxes in accordance. YOLO goes through its neural network while calculating errors for bounding boxes developed on the image of the best fit producing real-time results.

\emph{MTCNN}: The Multi-task Cascaded Convolutional Neural Networks (MTCNN) produces the best results on a collection of benchmark datasets among its contemporaries which can be made it to be very prevalent. Its capabilities of identifying the facial features like mouth and eyes are known as landmark detection.

The MTCNN utilizes a cascade assembly of three networks:
\begin{enumerate}
\item     P-Net or Proposal Network proposes a set of candidate facial sections after the image is resized while maintaining the aspect ratio into a series of different scales referred to as an image pyramid
 \item	R-Net or the Refine Network analyses the bounded boxes.
 \item	O-Net or the Output Network sets out facial landmark proposals.  
 \end{enumerate}

This model is known as a “Multi Task Network” since every model among the three in the cascade of (P, R and O Net) is trained over three tasks. These predictions include face classification with Landmark Localization and Bounding Container Regression. The outputs of the older phases are provided as inputs for the succeeding phases for the three models that are immediately unrelated. This paves for further processing to be done among the phases, like Non-Maximum Suppression (NMS) to be applied on the subject to filter the bounding containers projected by the foremost phase of Proposal-Net before giving them to the next phase of Refine-Network. The MTCNN algorithm is a three stage algorithm, it uses five landmark features on a face within an image to detect the face and create the bounding box for the face detected. Each of the stages passes the input to a CNN with the results steadily improving the result of the detection using the bounding boxes returned each time along with the scores and NMS.

\emph{YOLO}: Multiple boxes are bounded in every grid cells by the YOLO (You Only Look Once) prediction. For the true positive to have its loss determined the only cell with the maximum IoU (Intersection over the Union) is compared to the base truth. This model creates specialization that makes determining the boxes more efficient and reducing the loss every succeeding time. YOLO implements loss calculation method to determine the loss between the base reality of the image and the prediction of the model known as the ‘sum-squared error’.

\emph{Face Alignment using YOLO}: YOLO splits an image into a grid of thirteen by thirteen cells; each cell is answerable for predicting five bounding boxes. The bounding box outlines the parallelogram that surrounds an object. YOLOi uses a completely different approach. It applies one neural network to the total image. This network divides the image into regions and predicts bounding boxes and chances for every region. These bounding boxes area unit are weighted by the expected chances.

\emph{Wireless Sensor Networks}
\begin{itemize}
\item Sensor Node or Source Node gathers the environmental data and provides it to the Base station. 
\item A Sink or Base Station is in contact with the entire sensor node via some medium (Internet, Wi-Fi). It behaves like a gateway between the WSN and the outside world. It contains information from the entire connected sensor node.
\item Manager Node or User displays data for user. User can further analyze the data and query the sensor nodes. 
\end{itemize}

These participant node’s connections have some regular trends, such as 
\emph{Event detection}: Sensor node(s) must inform Base station(s) when they have data events corresponding to their tasks. This type of interaction requires more than one sensor node.
\emph{Function approximation}: An approximation mapping of the area specified in the application may be needed at the Base station. The sensor nodes are then used to approximate a position feature that points out changes in the physical value from one point to another.

A wireless sensor network consists of a gateway node being a server with higher computational capacity which receives the video feeds from the camera nodes. Server does the computing where it has the dataset of the faces. Classified and aligned faces are stored for the detection of faces in the video feed. Server can connect to multiple devices where it can receive and send data for monitoring. Multiple devices used for monitoring with a monitoring application can be used to authenticate and receive updates of the faces detected on the server. Information regarding blacklisting or any suspicious behavior can be received on monitoring application. 

\begin{figure} [htbp]
{
       \centering
       \includegraphics[width=0.8\linewidth]{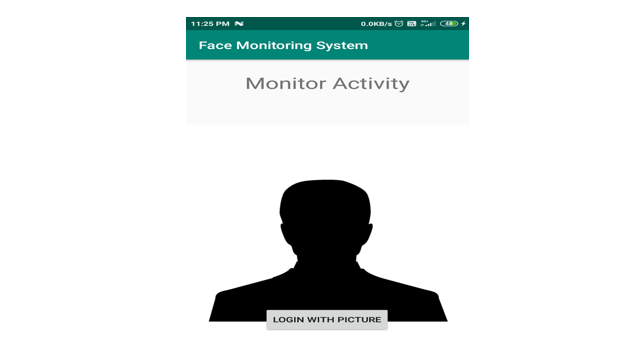}
       \caption{ Monitoring Application,login with face recognition}
}
       
The monitoring application can be accessed only by the individuals that have access registered in the database with the need to authenticate to receive the feed of the information produced by the nodes in the network. Figure 2 depicts the Face monitoring system application. Figure 3 shows the application receiving the information from the server, wherein the interval can be set and the faces identified can only be received over the period of set interval.

{
       \centering
       \includegraphics[width=0.6\textwidth]{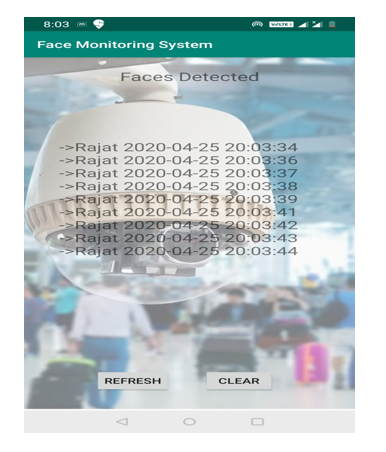}
       \caption{Information feed from gateway node}
}
\end{figure}
\section*{}
\section*{}
\section*{}
\section {Methodology}

High Level Design of the proposed method is depicted in figure 4. Dataset images are pre-processed, later segmentation is done to extract texture feature. Neural network classifier is applied to classify the images. 
\begin{figure}[h]
       \centering
       \includegraphics[width=0.6\textwidth]{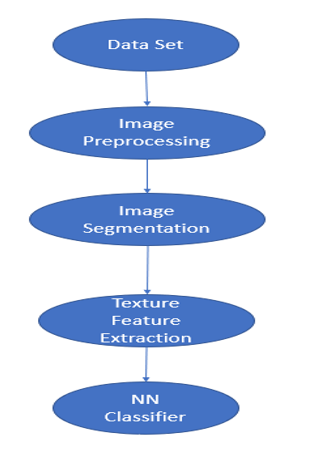}
       \caption{High Level Design}
\end{figure}

Figure 5 depicts the architecture of the proposed system. Video input from the cameras is sent to central server. Insertion or update of data is stored in the database.  

\begin{figure}[h]
       \centering
       \includegraphics[width=0.6\linewidth]{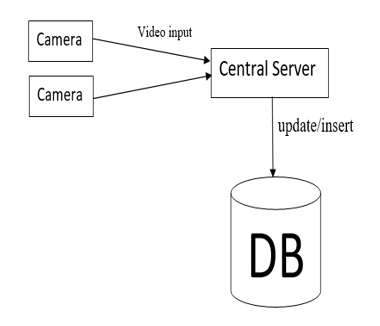}
       \caption{System Architecture}
\end{figure}

Simulation setup for the monitoring system are as follows
\begin{itemize}

\item Setting up a laptop to be a server owing to its high performance and it acts like base station. 
\item Another machine (laptop) with camera without the need or use of computation as a sensor node.
\item Designed an android app which interacts with the base station using a common wireless network.

\emph{Face detection and Face recognition}: 
\item Face detection and face recognition occurs at base station. 
\item Detect face using MTCNN algorithm. 
\item Detected faces are recognized by FaceNet. 
\item Initially classifier is set by providing the image set of people. 
\item Classifier is used by FaceNet to recognize the face.
\item A threshold value of 0.5 is set, below which the FaceNet model recognizes the person as a guest and stores his image with a generated id in the database for future reference.
\end{itemize}

\begin{figure}[h]
       \centering
       \includegraphics[width=0.9\textwidth]{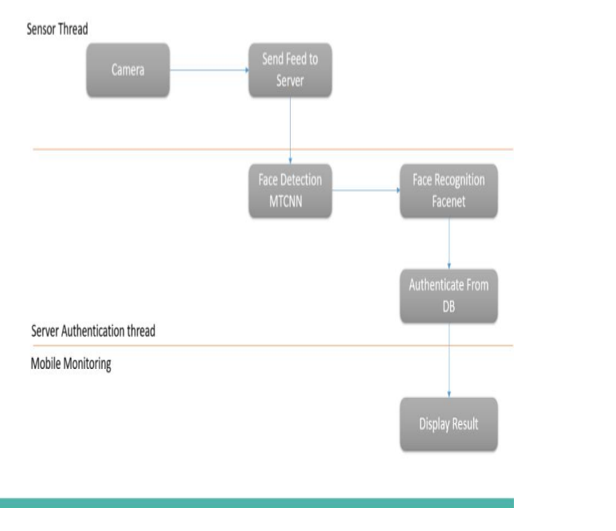}
       \caption{User Authentication}
\end{figure}
 
Figure 6 shows the authentication of user using MTCNN and FaceNet technique. Sensors capture the images and feeded to server. Face images are detected using MTCNN and faces are recognized using FaceNet. These face images are matched with stored templates in the database. Server authentication and mobile monitoring facilitate to display the results whether the user is authenticated or not.

\emph{WSN implementation}: A camera sensor node is connected to the base station using a Wi-Fi network. Wi-Fi is used as a common medium between the nodes and base station. All the connections between nodes and base stations is done through socket programming among several other methods \cite {17,18,19}. Camera nodes send its input to the base station. Base station runs the facial recognition scan and then further sends this information to all the end-user nodes (Android app) which are connected to the network.

\section{Results and Discussions} 

Performance evaluation based on the implementation are discussed in this section.

\emph{Training Images}: All the training images of face is represented in figure 7. The open source dataset of face images for the FaceNet, each subject containing varied numbers of images with FaceNet being able to recognize with minimum number of images for any given subject. 

\begin{figure}[h]
       \centering
       \includegraphics[width=1.0\textwidth]{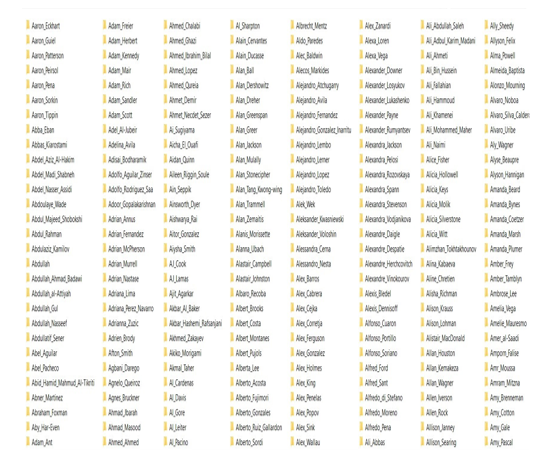}
       \caption{Dataset of training images }
\end{figure}
\section*{}
\section*{}
All the varied images of a particular subject name Abdullah Gul is shown in figure 8. The faces are detected in the source images and based on the features of faces detected; the bounding boxes are developed and images are cropped to obtain faces from the sources. Detected, aligned and cropped input faces pertaining to subject name Abdullah Gul is shown in figure 9.

\begin{figure}[h]
       \centering
       \includegraphics[width=1.0\textwidth]{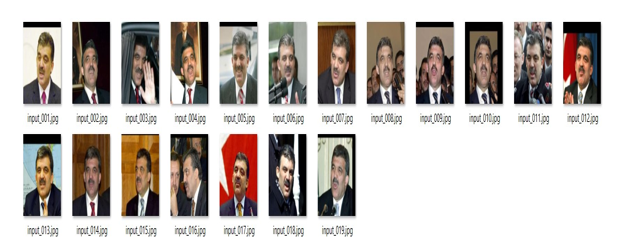}
       \caption{Varied images of a subject }
\end{figure}

\begin{figure}[h]
       \centering
       \includegraphics[width=1.0\textwidth]{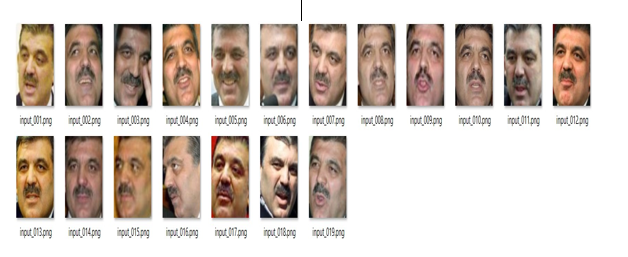}
       \caption{Detected and Cropped Faces }
\end{figure}

Figure 10 represents successfully scanned 13,233 images for creating the classifier and faces detected, aligned for FaceNet.

\begin{figure}[htbp]
{
    \centering
       \includegraphics[width=1.0\textwidth]{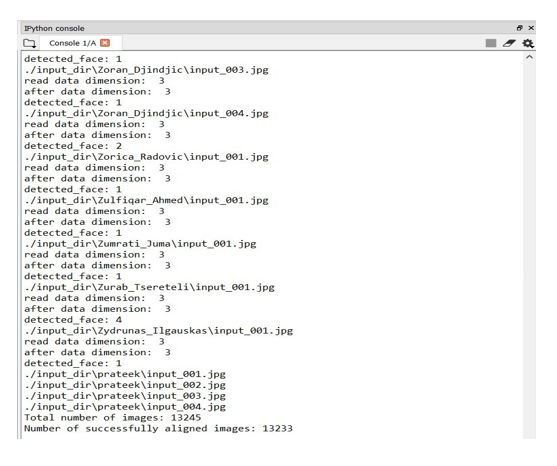}
       \caption{Faces detected and aligned for FaceNet}
}
Subject detection with 70 percent probability of face recognition is achieved. Figure 11 and 12 illustrates face recognition pertaining to subject name.

{      \centering
       \includegraphics[width=1.0\textwidth]{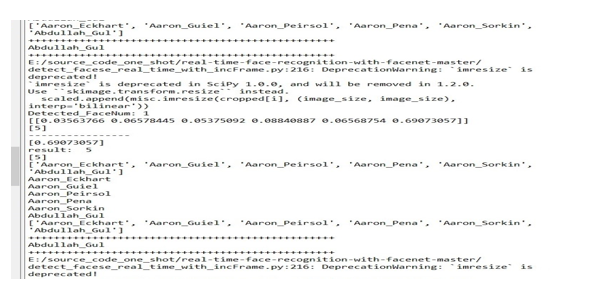}
       \caption{Face recognition of the subject}
}
\end{figure}

\begin{figure}[h]
      \centering
       \includegraphics[width=0.6\textwidth]{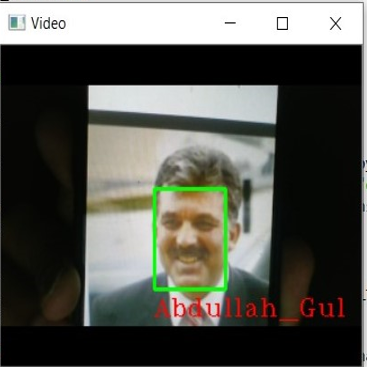}
       \caption{Face recognition to subject}
\end{figure}
\section*{}

The video feed displays bounding box as well as the name of the determined face along with recording of names and timestamp. The video feed can be monitored on the server real-time with complete information or the information passed to monitoring devices. The detected faces in the video feed from the nodes is also sent to monitoring applications as described in the figure 3 and 4, which gives information of the detected face as well the time when it was detected and provides with an alert when there is a suspicious or a flagged individual recognized in the video feed. 
\section*{}
\section*{}
\section*{}
\section{Scope for Future Work}

Recently, some group of researchers have released the new YOLOv2i framework, which leverages recent results in a deep learning network design to build a more efficient network, as well as use the anchor boxes idea from Faster-RCNNi to ease the learning problem for the network. The result is a detection system which is even better, achieving state-of-the-art performance at 78.6i mAPi on the Pascal VOCi detection dataset, while other systems, such as the improved version of Faster-RCNNi (Faster-RCNNi ResNet)i and SSD500,i only achieve 76.4i mAPi and 76.8i mAPi on the same test dataset. Maintaining a good Wi-Fi range in a facility is quite challenging, as all nodes need to be connected to the network. The research can further be taken a step ahead by using more than one routers to amplify the network coverage and sync all the nodes connected to the network more seamlessly.

\section{Conclusion}

Wireless Sensor Networks provide a way to automate authentication, security and database management. As one node acting as master or server node connected to database, is capable to perform heavy tasks. Server node is connected to various other slave or client nodes which depend on server node for the data. The topology provides a way in which the client nodes are not required to have high computational power. Client nodes must have a good connection with the server. It’s a challenging task to keep a good connection with the server since it depends on the network connectivity and stability. The proposed work have used Wi-Fi as a medium. A common Wi-Fi network between all the nodes is an effective way of communication, enabling our solution a self-sustainable network which does not depend on Internet. Servers, nodes of cameras and monitoring devices offer smooth security, as one can easily monitor the information obtained and analyzed from the feed of cameras. This can be easily implemented by choosing the best of the sensor nodes.
Proposed work recognizes faces of user’s choice. Face recognition with FaceNet also has a drawback where in it always locates the closest image or face based on the Euclidean distance – Triplet loss. The drawback of the FaceNet can also be overcome by setting a properly determined threshold value and a new id can be given to a new unknown face for recording it to the database. 

\bibliographystyle{unsrt}
\bibliography{mybib}

\end{document}